Microheterogeneity and Single-Molecule Mixture Science: An Introduction

# Microheterogeneity and Single-Molecule Mixture Science: An Introduction

Yu Tang*

**Abstract**: Microheterogeneity, as a fundamental natural property, is widely presented in a range of microscopic systems such as polymer systems, biomacromolecular systems, and nanosystems, however, the construction of molecular systems of this form has not yet been realized at the level of organic molecules with well-defined structural compositions. Inspired by the "chemical space" concept, I recently reported a study of the single-molecule mixture state, In this preprint, a detailed discussion of microheterogeneity and the concept of "single-molecule mixture" has been provided.

## 1. Background

There is a famous saying that "no two snowflakes in the world are the same", which reveals a fundamental law of nature: no two objects in nature are identical. In the microscopic world, this law also exists, hence the term “microheterogeneity” to describe this phenomenon.

However, in the molecular world we are currently studying, the situation is different. If we compare a molecule to a snowflake, we find that a pure organic molecule is composed of a single molecule with the same structural formula, i.e., from the point of view of the structural formula of the composition, every snowflake is the same! (Of course, molecules are always in motion, and if we consider conformation, each molecule is different.)

It is natural to wonder whether there is a class of substances in which each molecule has a different structural formula. In other words, are there substances in the form of "unimolecular mixtures"?

This idea, at first glance, sounds impossible and even absurd. After all, the number of molecules in one mole is $6.02\times10^{23}$, and the number of molecules that have been synthesized is on the order of $10^8$, and it is simply not possible to make a mixture of one molecule from each molecule that has been synthesized. Therefore, there does not seem to be a feasible way to obtain such a conceivable "single-molecule mixture".
However, from another point of view, if we can make full use of the principle of permutation and combination, it is possible to obtain molecules with different structural formulas by exponential expansion, and thus obtain a unique form of "single-molecule mixtures". This idea came to my mind in April 2018, and was published on arXiv in 2021 (Yu Tang, Single Molecule Mixture: A Concept in Polymer Science, arXiv:2102.05845, https://doi.org/ 10.48550/arXiv.2102.05845). In a series of subsequent sections, I will discuss the concept of "Single-Molecule Mixture" in detail.

## 2. The basic principle of single-molecule mixture system construction

The basic principle of the construction of single-molecule mixture system is: firstly, use the principle of permutation and combination to construct a theoretical molecular structure space containing a sufficiently large number of molecules, and then use the approximate equal probability of random synthesis to synthesise the molecules in this space, if the number of molecules in the actual synthesised samples is much less than the theoretical value, then we may obtain a "single-molecule mixture" sample with a large probability.

There are three important parameters in a single-molecular mixture system: the total number of isomers **n** in the structural space to which the product belongs, the total number of molecules **m** of the product obtained according to the equiprobable stochastic synthesis method, and the probability that the resulting product will be in the single-molecular mixture state **P**. How is the relationship between them calculated? It is illustrated here by a mathematical model as shown in Fig. 1.

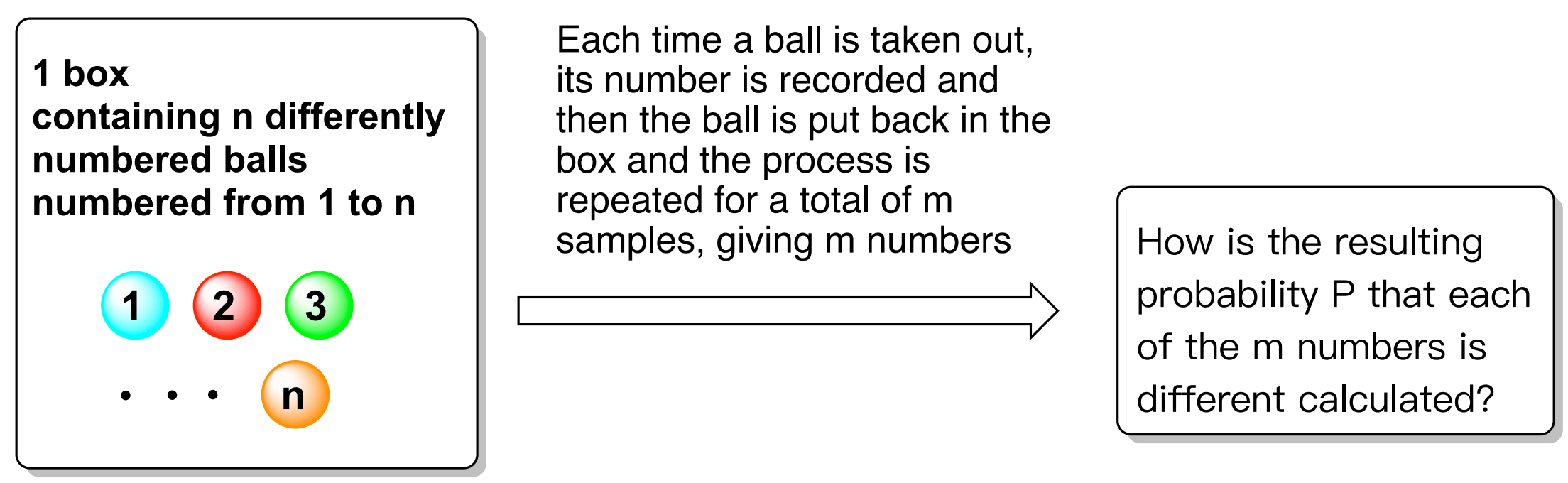

**Fig. 1**

Let us consider this model: 1 box containing **n** differently numbered balls, numbered from 1 to **n**. Each time a ball is removed, its number is recorded, and then the ball is put back into the box, and the sampling process is repeated for a total of **m** times, resulting in **m** numbers, and what is the probability **P** that the resulting numbers are each different?

This is a typical probability calculation problem, and **P** is calculated as follows:

$$P = \frac{A_n^m}{n^m} = \frac{n(n-1)(n-2)\cdots\cdots(n-m+1)}{n^m}$$

$$= (1-\frac{1}{n})(1-\frac{2}{n})\cdots\cdots(1-\frac{m-1}{n})$$

$$> (1-\frac{m-1}{n})^{m-1} \qquad \textbf{eq.1}$$

This equation satisfies the conditions for the establishment of Bernoulli's inequality, so it can be approximated by Bernoulli's inequality, which is obtained:

$$P > 1-\frac{(m-1)^2}{n} \qquad \textbf{eq.2}$$

In this way, the relational equation between **m**, **n** and **P** is obtained. Through this relational equation, the following three important information can be calculated:

1) The minimum value of **P** for a given number **n** of spatial samples and a given

number **m** of samples;

2) The minimum value of the number of spatial samples **n** required to make **P** greater than a certain value for a given number of samples **m**;

3) for a given total number of spatial samples **n**, the maximum value of the number of samples **m** to make **P** greater than a certain value;

This mathematical model is a good solution to the computational problems associated with single-molecule mixture systems. In the single-molecule mixture model, the total number of isomers in a structural space corresponds to **n** in the above model, the number of molecules in the sample obtained by the equal probability random synthesis method corresponds to **m** in the above model, and the probability that the product exists in a single-molecule mixture state corresponds to **P** in the above model, and the relationship between the three is the same as that in the above model eq.2.

According to the above formula, we can get the following results: for a structure space containing **n** isomers, 1 mol ($6.022\times10^{23}$) molecules of this space are synthesized by equal probability random synthesis method, and the probability that the resulting product exists in a single molecule mixed state is **P**. The minimum number of isomers **n** of this space is $3.63\times10^{50}$ for $P > 0.999$, and the minimum number of **n** should be $3.63 \times 10^{52}$ for $P > 0.99999$, as shown in Fig. 2.

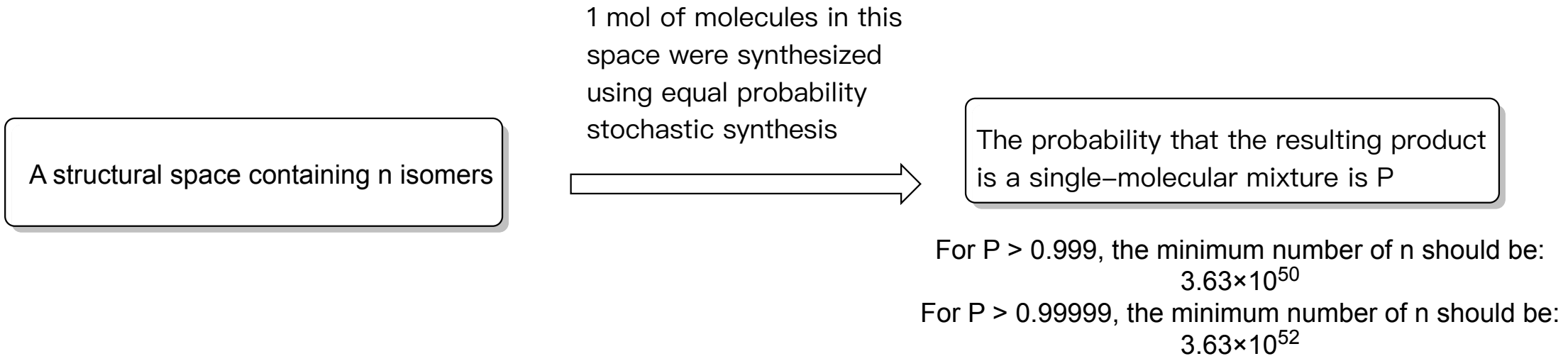

**Fig. 2**

## 3. Sugar chains, the ideal backbone for building single-molecule mixture models

A typical process for constructing a single-molecule mixture model is shown in Fig. 3, in which firstly, the basic modular units are designed and prepared, and then these modular units are chemically spliced to obtain a single-molecule mixture model system. In this, the design of the basic modular units and their splicing methods are crucial.

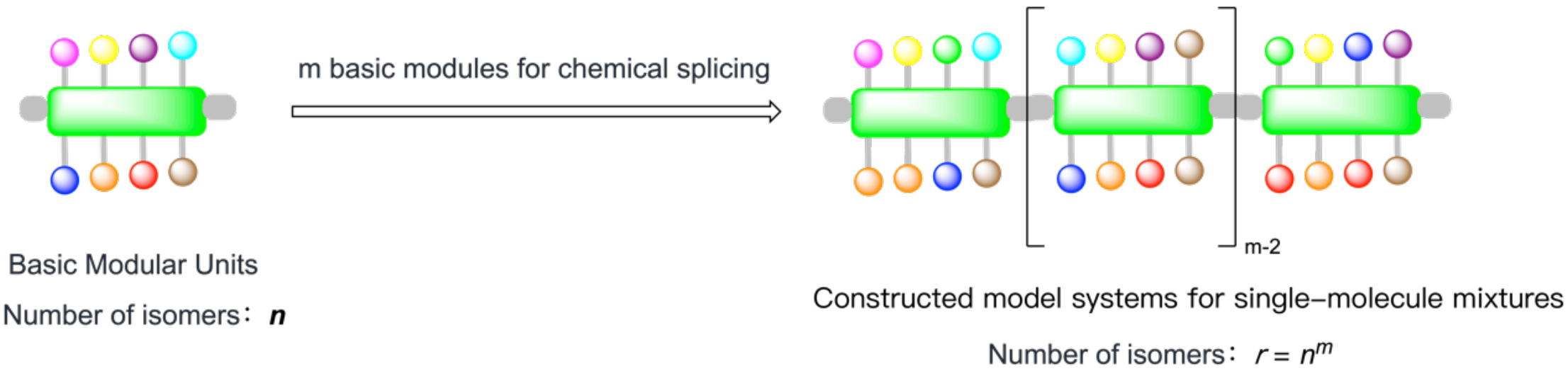

**Fig. 3**

In the previous discussion[1], the basic modular unit I designed was substituted mannitol, and the splicing method was etherification, the biggest problem this system may face in the actual synthesis is that the efficiency of etherification reaction is not high enough, which may affect the purity of the product. Thus, it is necessary to consider other modules and splicing methods.

In my recent work, I have further considered the feasibility of adopting sugar building blocks as the basic modules and chemical glycosylation method as the splicing method, and after comprehensive consideration, I have found that the sugar chains are the ideal skeletal materials for constructing the single-molecular mixture model. The main reasons for this are: 1) the synthesis method of monosaccharide building blocks with various types of protecting groups is more mature; 2) the technology of splicing monosaccharide building blocks using chemical glycosylation method is more mature and reliable; 3) the purification and characterization techniques of the synthesised substituted sugar chains are more mature and reliable. Thus, glycosynthesis technology will play a key role in the development of single-molecule mixture science.

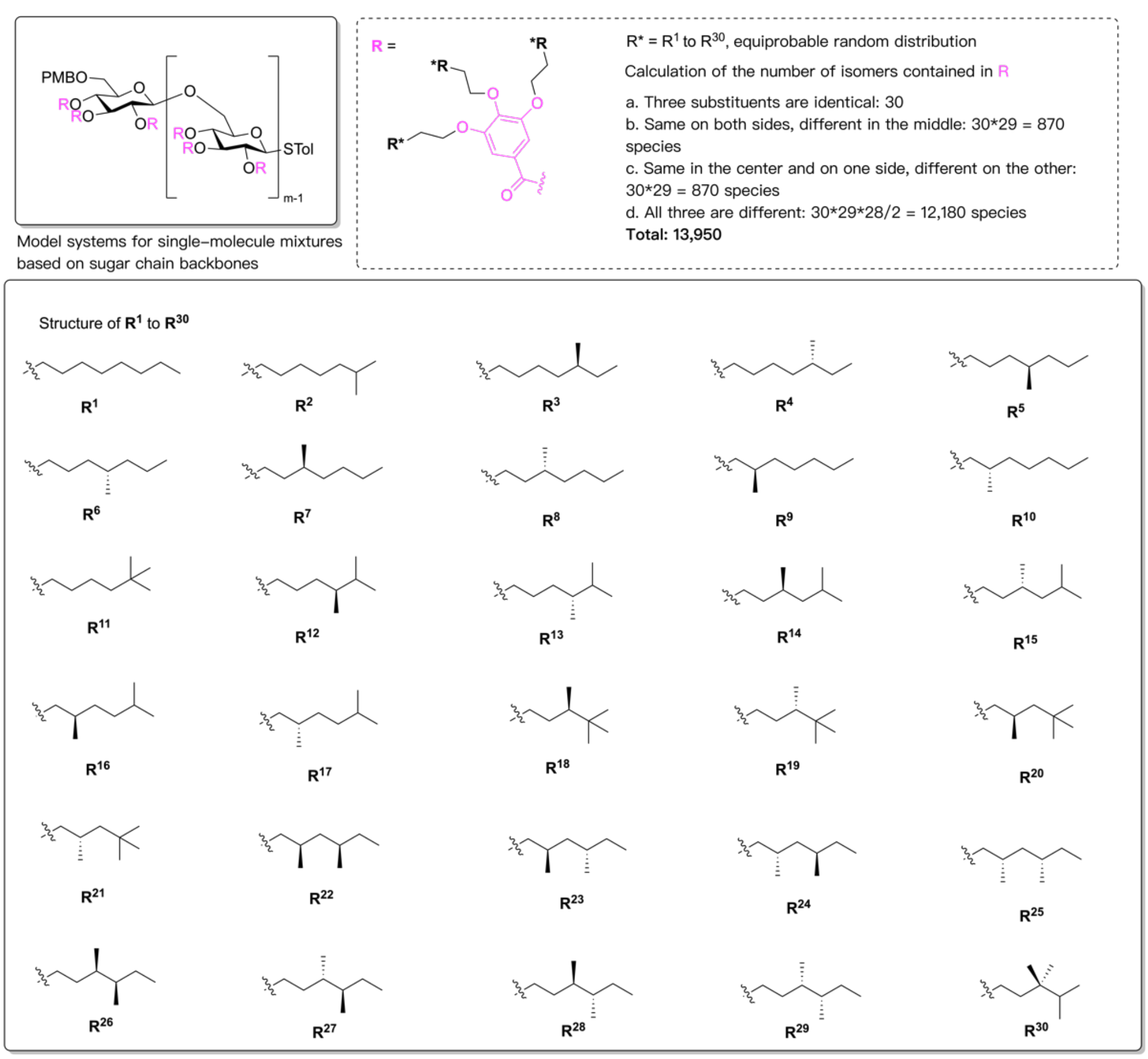

**Fig. 4**

In the following, I illustrate the construction of a single-molecule mixture system based on a sugar chain backbone with a specific example.

The constructed single-molecule mixture model is shown in Fig. 4, whose basic skeleton consists of polydextrose chains, in which three different substituents R can be attached to each sugar, and the number of isomers for each R is calculated to be 13,950, with an equal probability random distribution. (Since the structural differences of the substituents are extremely small, the different substituents and the numerous number of isomers in the synthesis and splicing of these sugar blocks have little effect on the reaction and purification process, and the experiments are essentially the same as when pure material is used as raw material.)

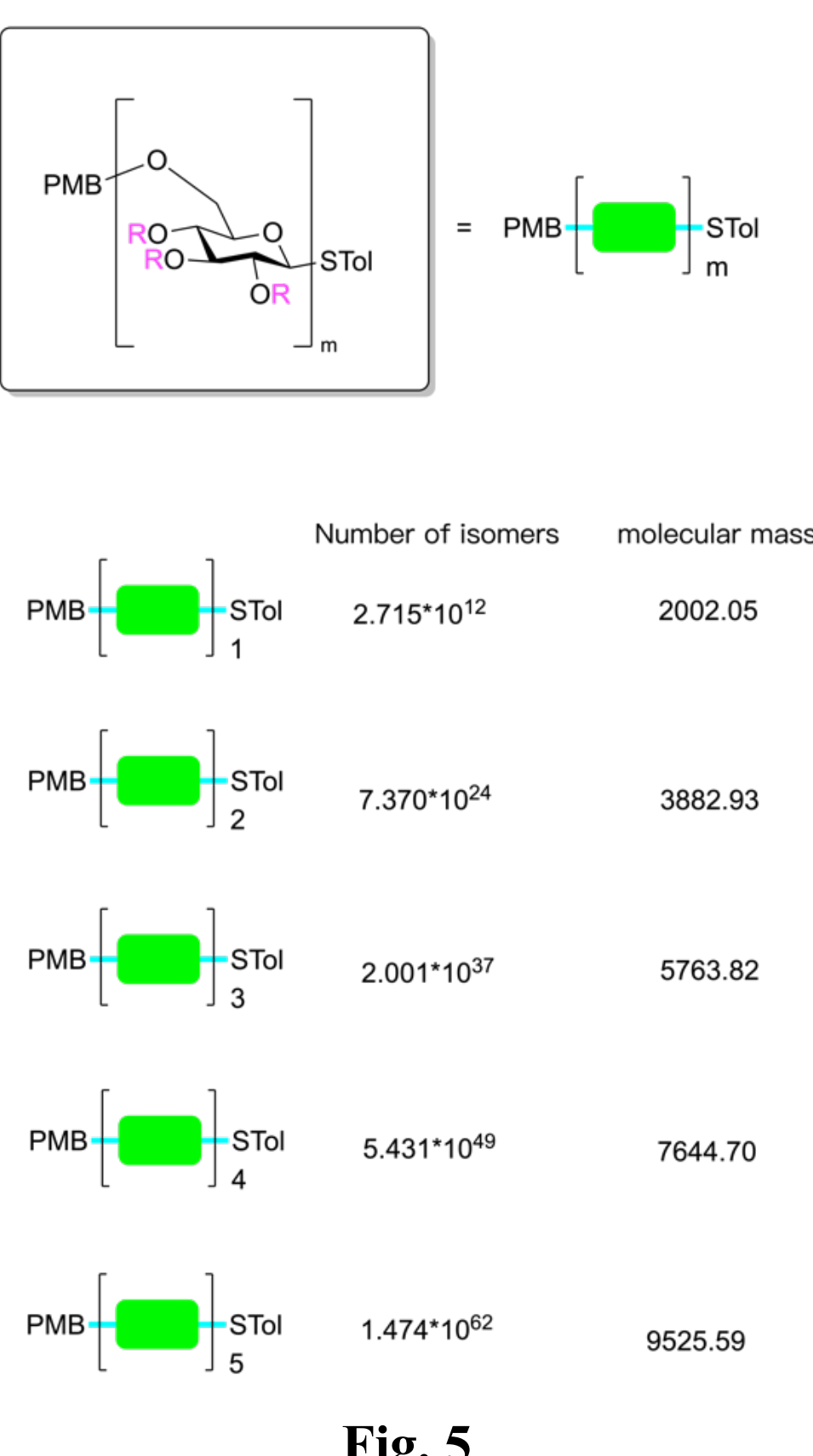

**Fig. 5**

For sugar chains with different numbers of polymerisation (m), the molecular weights and the number of isomers are shown in Fig. 5. As can be seen from the figure, when m = 4, the number of isomers of the system is already as high as $5.431\times10^{49}$ species, which, according to calculations, is a number, relative to the usual scale of preparation (no more than 1 mmol), sufficient to ensure that the product is distributed in a single-molecule mixture state.

From this example, it can be seen that the sugar chain is an ideal backbone material in the construction of single-molecule mixture systems, and the knowledge in glycoscience will play an important role in the construction of single-molecule mixture systems, and glycoscientists are facing a broad development opportunity in the field of single-molecule mixture research!

## 4. Construction of a spherical single-molecule mixture model

The single-molecule mixture model can be constructed using not only chain-like molecular skeletons, but also globular molecular skeletons, and its chemical splicing can be performed not only by etherification reactions, chemical glycosylation reactions, but also by click reactions. In the previous examples, I mainly showed the chain-like single-molecule mixture model, and here, I provide an example of a globular single-molecule mixture model, which belongs to the structural space **A1** defined as shown in Fig. 6.

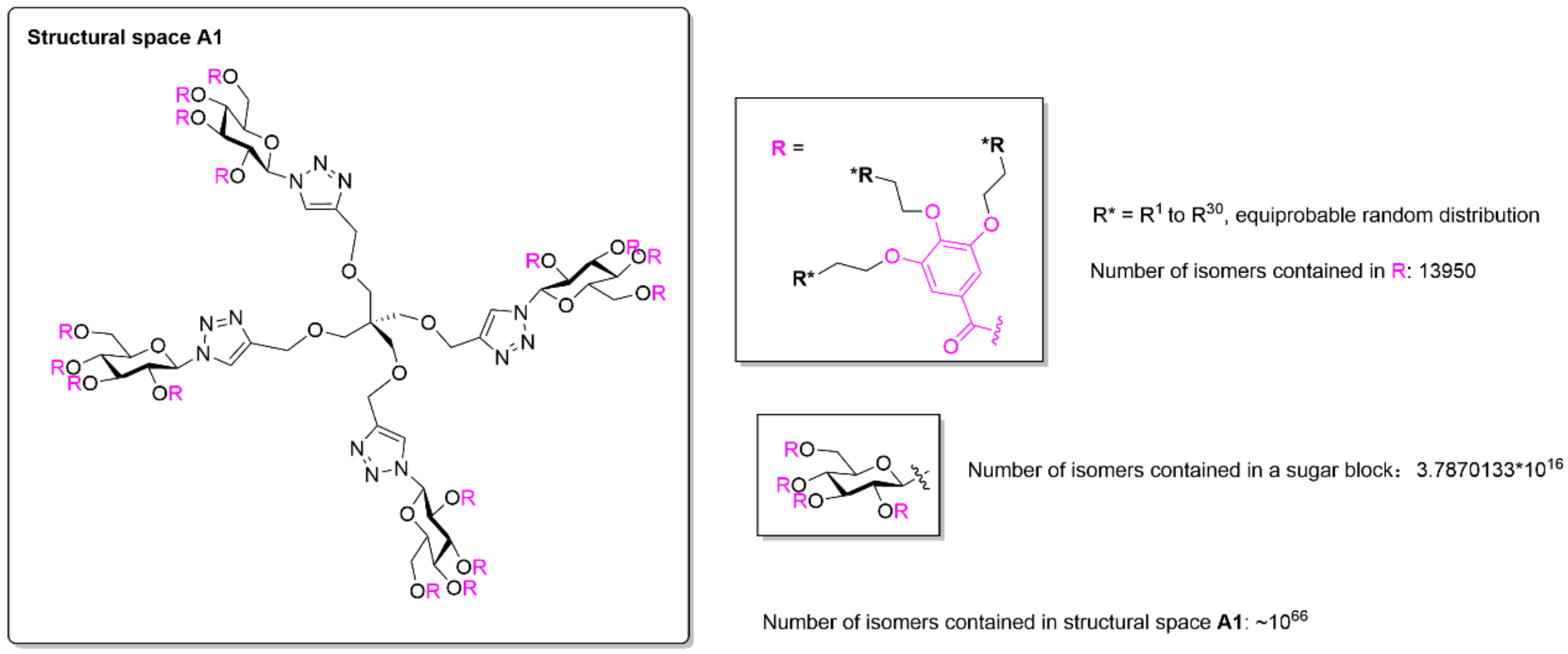

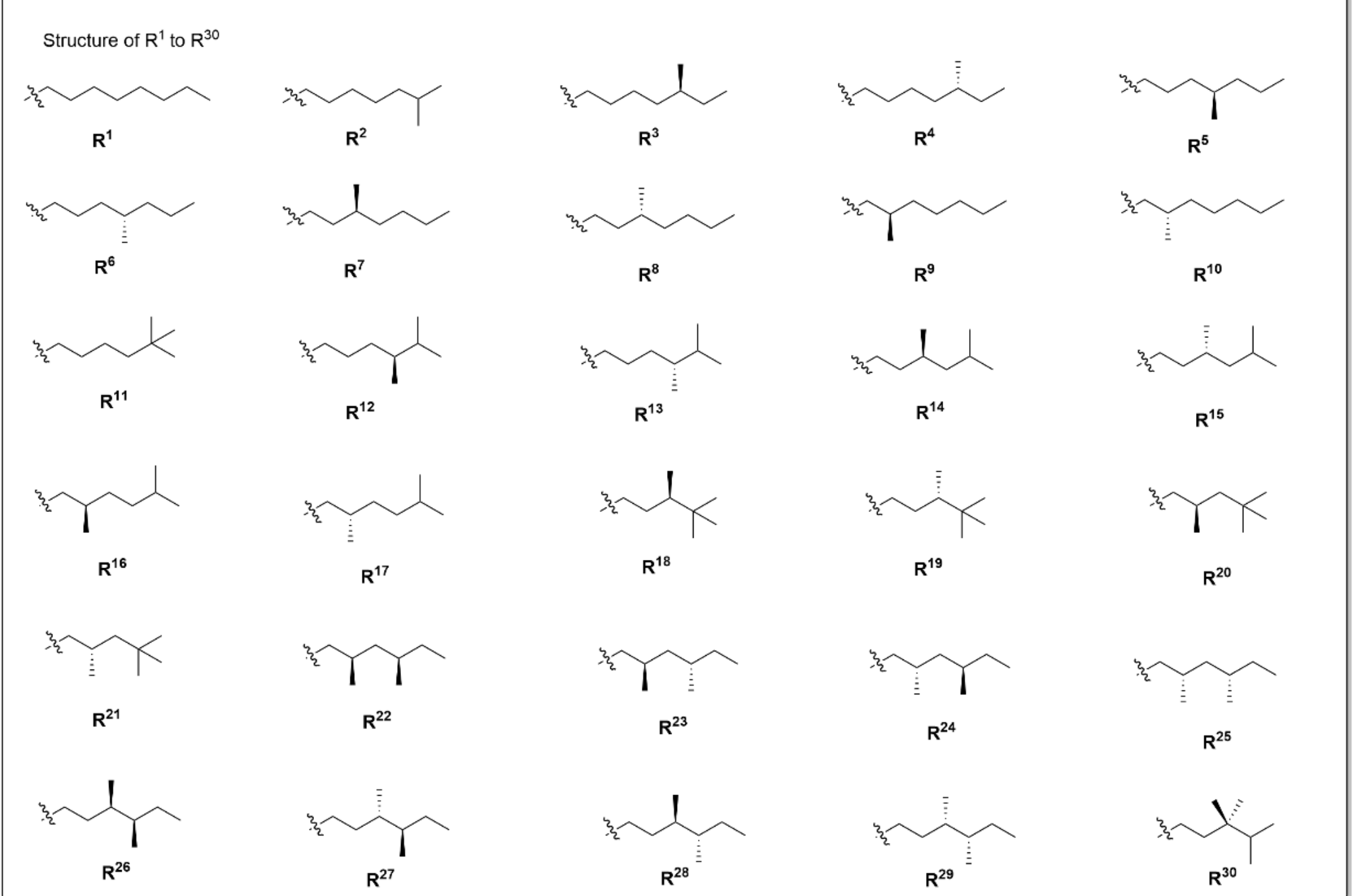

**Fig. 6**

Each molecule in this space has an equal molecular weight, and the total number of isomers is on the order of $10^{66}$, which is sufficient to ensure that the product is in a single-molecule mixture state at the usual scale of preparation (<<1 mol). The preparation was mainly linked using a click reaction as shown in Figure 7, which is well established and convenient for synthesis and characterization.

**Fig. 7**

The spherical single-molecule mixture model constructed here may present different and exotic properties relative to the single alkyl-substituted purity model, which is yet to be explored in detail. In the near future, I will conduct relevant experimental explorations, which will then unravel the mystery of this class of molecules.

## 5. A cyclic single-molecule mixture model based on the cyclodextrin backbone

With β-cyclodextrin as the backbone, it is easy to construct a single-molecule mixture model of cyclic structure by taking advantage of the fact that it can easily undergo fully acylation or fully alkylation reactions, as shown in Fig. 13. The alkyl or acyl portion thereof has an extremely rich design space. The resulting single-molecule mixtures can be conveniently purified and analyzed by existing separation and purification and structural characterization techniques. The highlight of this synthetic scheme is the great simplicity, which allows the construction of a complex cyclic structure of the single-molecule mixture system in only one step reaction.

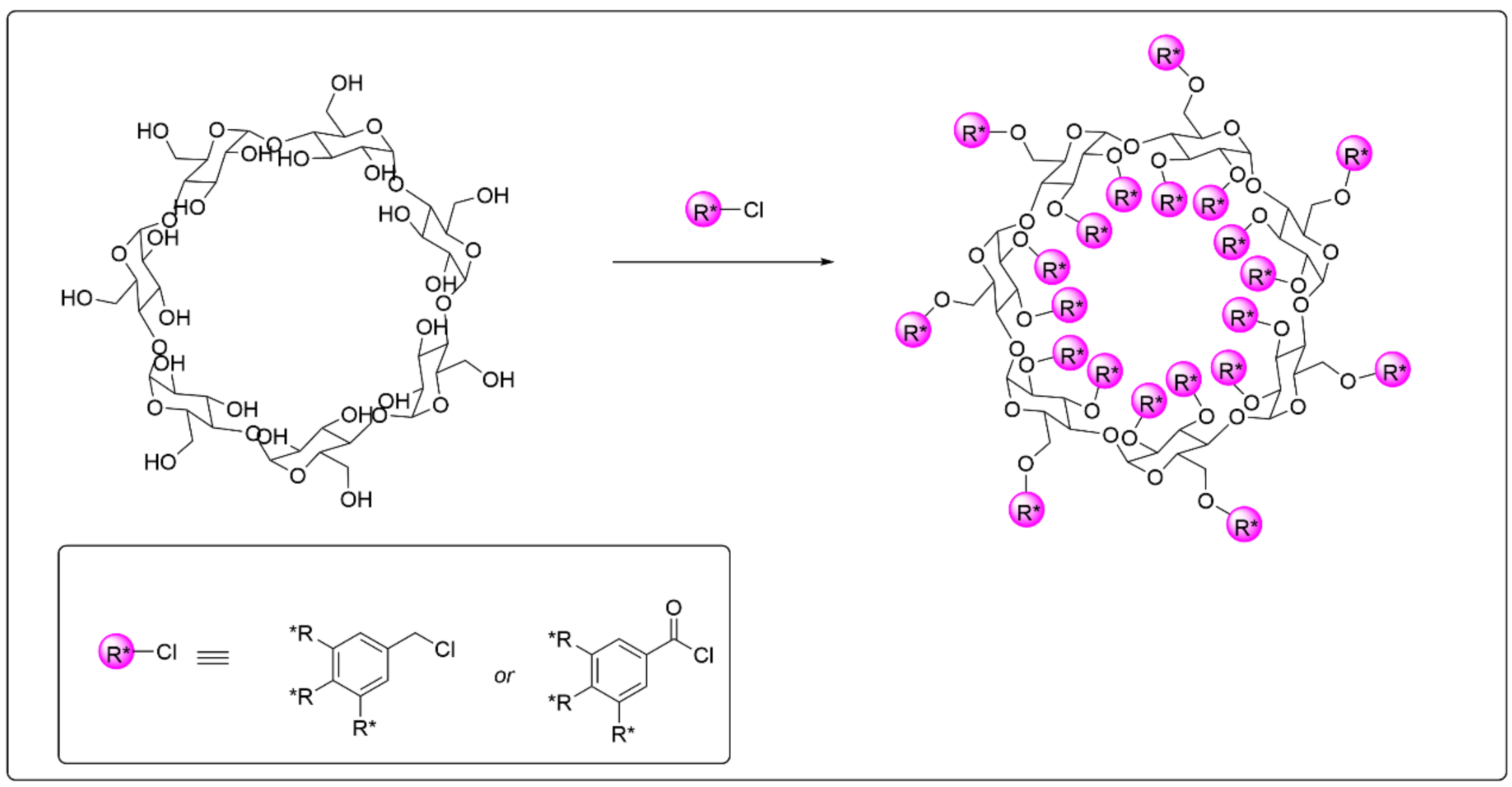

**Fig. 8 Construction of a cyclic single-molecule mixture model based on the β-**

**cyclodextrin backbone**

## 6. A class of single-molecule mixture models based on the monosaccharide backbone

Designing single-molecule mixture systems with more concise structures and easier preparation is a worthy and continuous goal in the field of single-molecule mixture science. On the basis of the previous series of designs, I have further designed a class of more concise single-molecule mixture models based on the monosaccharide backbone, as shown in Fig. 9. The system used α- glucose methyl glycoside as the backbone and 3,4,5-tris[3,4,5-tris(decyl)benzyloxy]benzoyl as the modifying group, in which the decyl substituents were 30 structures with equal probability random distribution as shown in Fig.1

Structural space A

Molecular weight: 7509.7870

Calculation of the total number of isomers in the structural space A:

a. The number of isomers in a unit: 13950

b. The number of isomers of R

$13950^3 = 2.71\times10^{12}$

c. The total number of isomers in that structural space:

$r = (13950^3)^4 = 5.43\times10^{49}$

Structure of R*

Fig. 9

According to the calculations, the molecular weight of the system was 7509.7870,

and the number of the total isomers in the structural space was $5.43 \times 10^{49}$, in the presence of 1 mmol ( 7.5 g) preparation scale, the probability of the product presenting a single molecule mixed state was > 0.999. Total acylation of hydroxyl groups on monosaccharides is an easily achievable process, and thus, the preparation of this system is more facile compared to the previous systems.

## 7. A Scheme for Constructing Single-Molecule Mixture Models Using Deuterium Labeling Methods

In the single-molecule mixture models I constructed above, all of them used the method of changing the structure of the alkyl side chain to create different isomers. Although this strategy is very effective, there is a potential problem that isomers with different alkyl side chains belong to different molecules with different structures after all, and there is still uncertainty as to whether the mixtures formed by them can be effectively separated and purified by existing purification methods. Here I considered another scheme to construct a single-molecule mixture model using the deuterium labeling method, the basic principle of which is shown in Fig. 10.

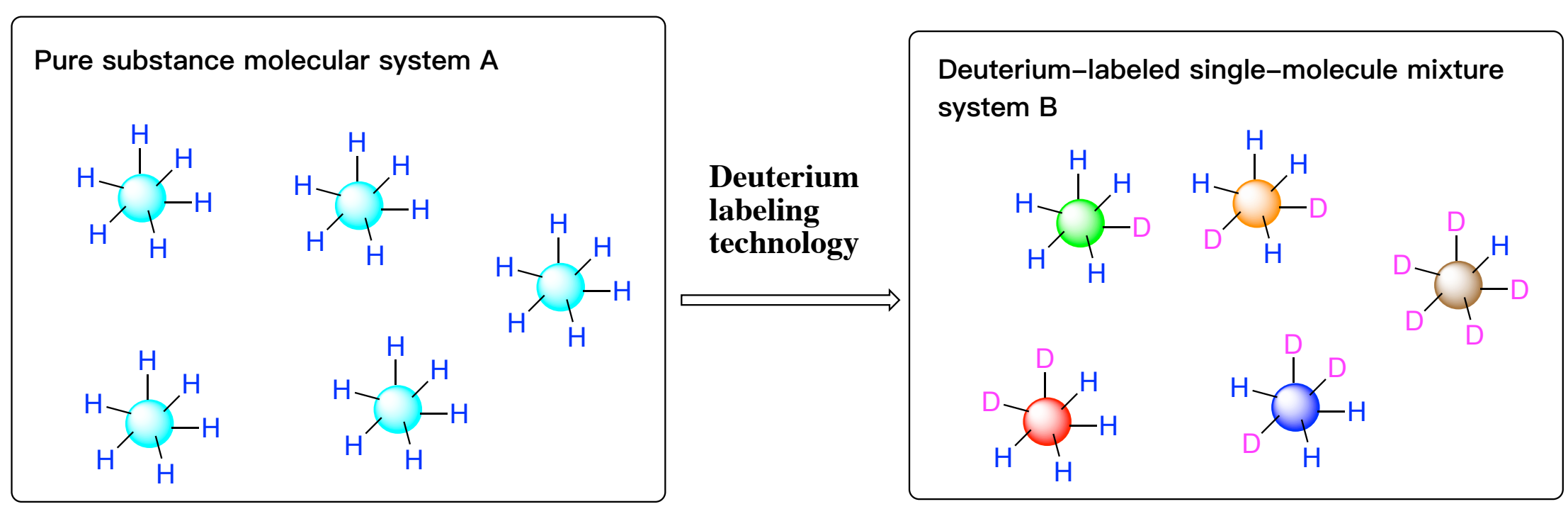

**Fig. 10**

Taking a pure substance molecular system **A** containing many alkyl hydrogen substitution sites in its structure as a prototype, a deuterium-labeled single-molecule mixture system was created by deuterium-labeling the hydrogen atoms in it, in which the structure of each molecule differed in the deuterium-labeled sites and number.

Based on this principle, I constructed a deuterium-labeled single-molecule mixture system based on a monosaccharide backbone, as shown in Fig. 11.

This system is based on the glucose monosaccharide backbone, and by introducing 30 different deuterium-labeled *n*-decyl substituent side chains into the molecule a structural space containing a sufficiently large number of deuterium-labeled isomers is created. The molecular weights of the deuterium-labeled isomeric structures in this space are distributed in an interval, with a minimum molecular weight $MW_{min}$ of 7504.1372 (corresponding to the molecular formula $C_{479}H_{822}O_{58}$) for non-deuterated substituents and a maximum molecular weight of 7612.8151 (corresponding to the molecular formula $C_{479}H_{714}D_{108}O_{58}$) for the maximum theoretical deuterium substitution rate.

During the synthesis of molecules in this structural space, since the difference between different isomers only lies in the deuterium substitution rate and deuterium substitution site, the physical properties of the mixtures composed of each isomer have very little difference, and they can be effectively separated and purified by the usual means of separation and purification. The difficulties that may be encountered in the preparation of single-molecule mixtures are completely overcome, and the experimental feasibility is high.

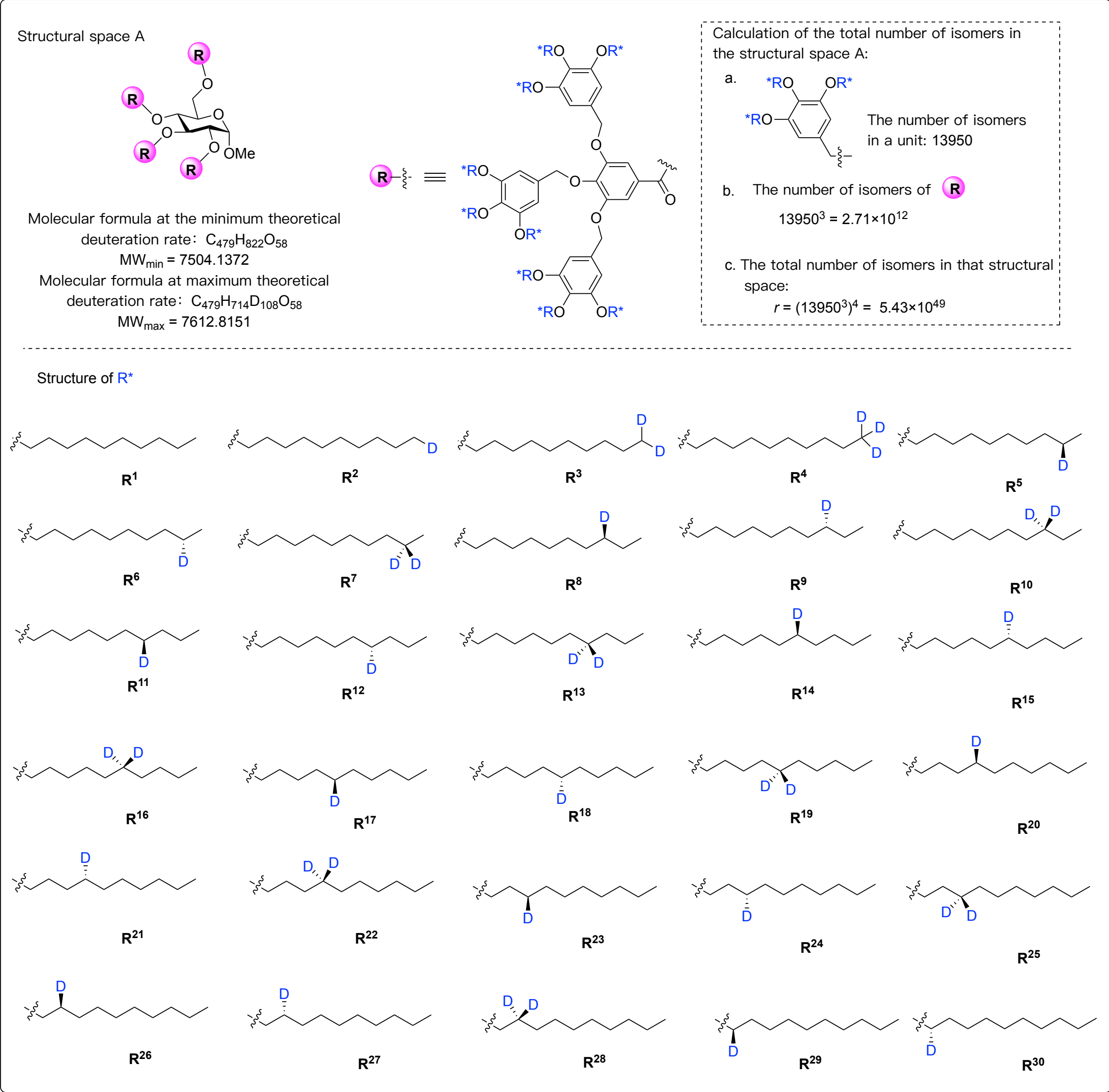

Fig. 11

In the prepared product samples, the deuterium substitution rate distribution and molecular weight distribution can be accurately simulated by theoretical calculations, and by comparing this simulation result with the mass spectrometry measurement result, it can be verified whether the product presents the theoretical distribution of the single-molecule mixtures, which provides an effective method to verify the single-molecule mixtures through the combination of theoretical simulation and actual measurement.

In summary, the single-molecule mixture model constructed by deuterium labeling method is a very feasible and valuable method for preparing single-molecule mixture models, and in the future, this method will play a very crucial role in the development of single-molecule mixture science.

## 8. A scheme for constructing single-molecule mixture models using a fluorine labeling strategy

By first constructing a basic structural skeleton and then labeling the groups in it, it is a very effective way to construct single-molecule mixture models. In the previous discussion, I mainly used alkyl labeling and hydrogen isotope labeling to construct single-molecule mixture models, and here, I provide another scheme for constructing single-molecule mixture models using a fluorine labeling strategy.

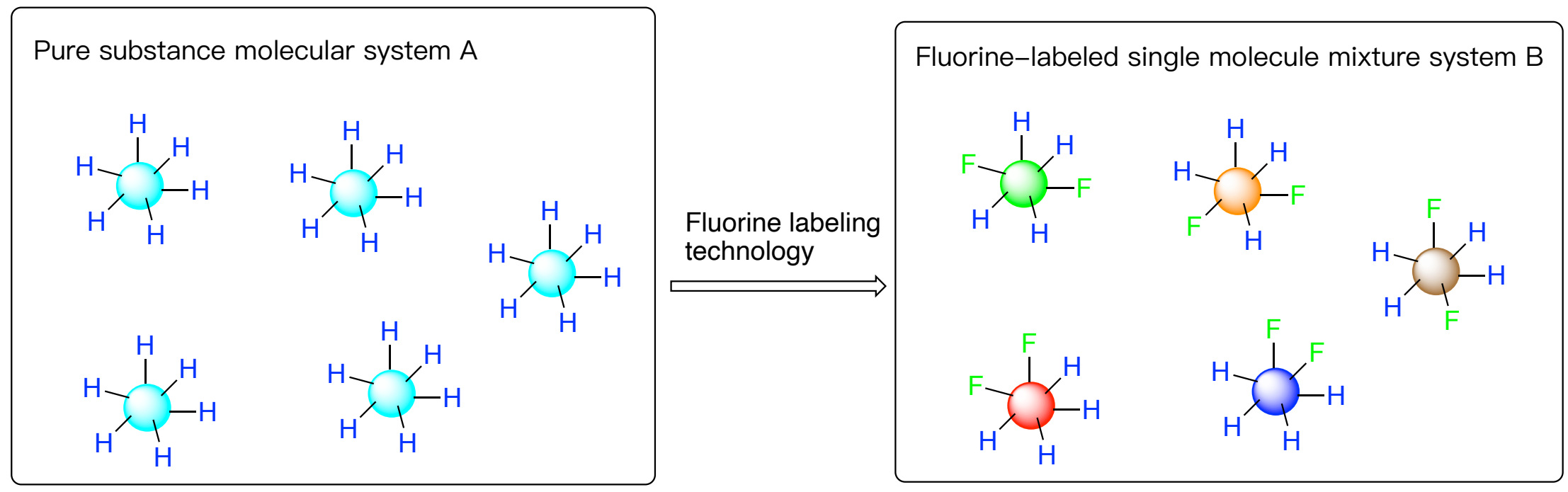

Fig. 12

The principle of this strategy is shown in Figure 12. Taking a pure substance molecular system A, which contains many alkyl hydrogen substitution sites in its structure, as a prototype, a fluorine-labeled single-molecule mixture system is created by fluorine-substituted labeling of hydrogen atoms in it, in which the structure of each molecule differs in the sites of the fluorine substituents.

Based on this principle, I constructed a fluorine-labeled single-molecule mixture system based on the fibrous disaccharide backbone, as shown in Figure 13.

In this system, a structural space containing a sufficiently large number of fluorine-labeled isomers was created by introducing 19 different fluorine-labeled n-dodecyl substituent side chains into the molecule. The number of isomers in this structural space is 5.10095204 × 1074, which is sufficient to ensure that the product exists in a single-molecule mixture state at the usual scale of preparation (1 mol level). Each molecule in this space contains an equal number of fluorine atoms (63) in its structure and has an equal molecular weight of 16059.43 (corresponding to the molecular formula $C_{965}H_{1627}F_{63}O_{102}$).

During the synthesis of the molecules in this structural space, since the difference between different isomers only lies in the difference of the fluorine substitution sites, the physical properties of the mixtures composed of the isomers are less different, and it is likely to be able to be effectively separated and purified by the usual separation and purification means, with a high experimental feasibility.

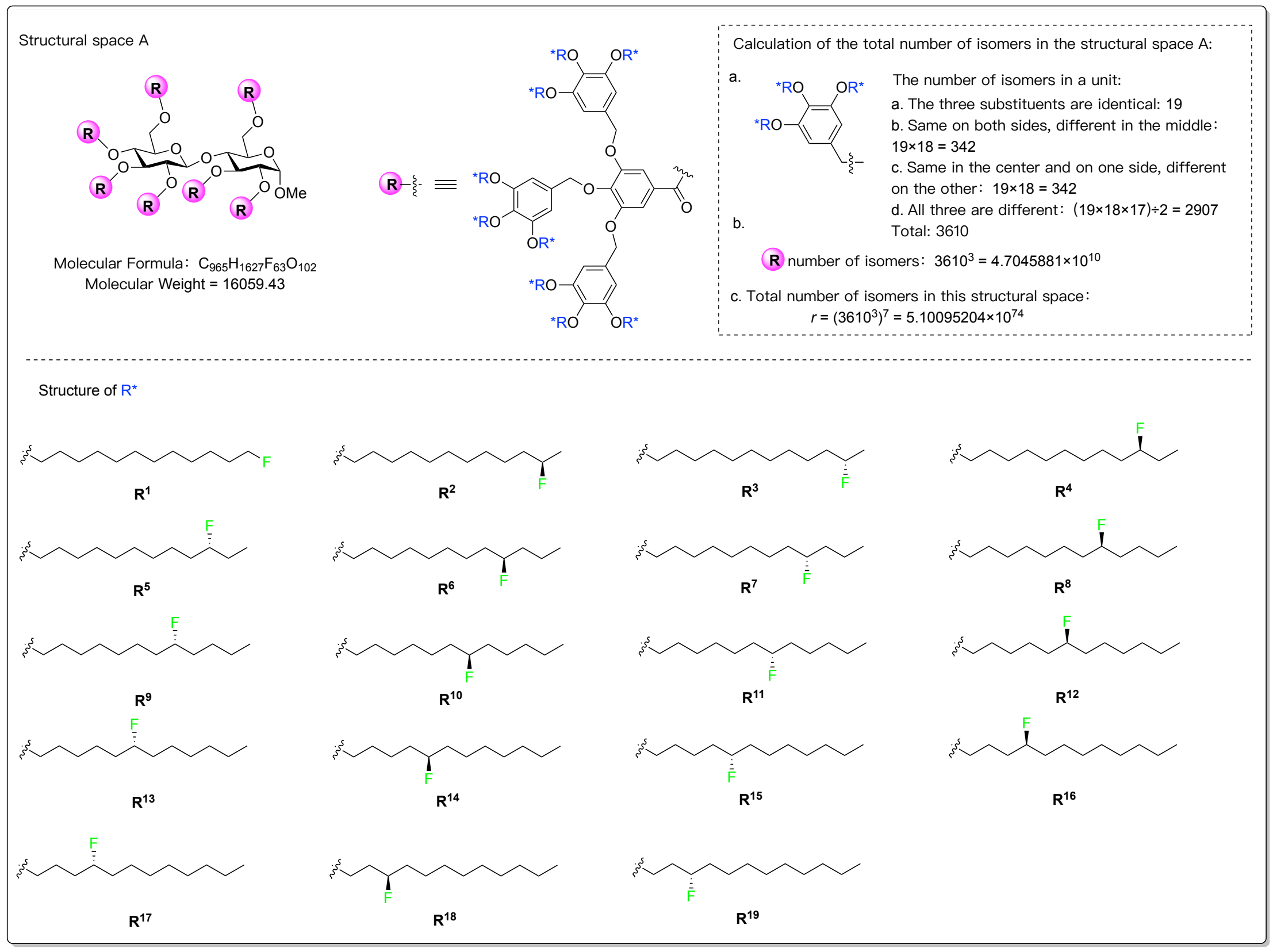

Fig. 13

In summary, the use of fluorine labeling strategy to construct single-molecule mixture models is a feasible and valuable method to prepare single-molecule mixture models, and this method will play an important role in the development of single-molecule mixture science in the future.

## 9. Understanding the architecture of single-molecule mixtures

The single-molecule mixtures prepared according to the scheme we devised in section 3-5 have three distinctive features:

1) The molecular weight of each molecule is equal;

2) The type of substituent group in each substitution site in the molecule is fixed and randomly distributed with equal probability;

3) The structure corresponding to each molecule in the sample belongs to the same structural space, and the theoretical number of isomers in this structural space can be calculated in a reasonable way.

So, how can we further understand the structure of this single-molecule mixture system?

Obviously, this single-molecule mixture system is not a disordered system, and they have their intrinsic laws and characteristics.

In the following, we use the method of model analysis to further discuss the structure of this type of single-molecule mixtures.

In the chain-like single-molecule mixture system shown in Fig. 14, the molecule contains m substitution sites, each of which is replaced by a substituent R, where R = $R^{A1}$, $R^{A2}$, $R^{A3}$.......$R^{An}$, which are randomly distributed with equal probability.

Borrowing the idea of orbital hybridization from molecular orbital theory, we define a new substituent $R^{A*}$ as a structural hybrid of $R^{A1}$, $R^{A2}$, $R^{A3}$.......$R^{An}$, then in this model each substitution site is replaced by a single substituent $R^{A*}$. This monomolecular mixture system can be regarded as a new form of "pure substance", and it can be expected that some of the basic physicochemical properties of this sample of monomolecular mixtures have constant values as those of pure substances in the usual sense.

**Fig. 14**

From the above analysis, it can be seen that we can better understand the structure of single-molecule mixtures by adopting a hybridization approach at the molecular structure level, drawing on the classical theory of orbital hybridization. The reason this idea is feasible is that the number of molecules as well as the number of substitution sites in a macroscopic sample are large values, and the use of statistical averaging is perfectly feasible.

Next, let us look at a concrete example.

Shown in Fig. 15 is a specific example of a single molecule mixture system that we constructed. This system is a 24-aggregate of substituted mannitol, and in the side chain of the molecule, each free alcohol hydroxyl group is randomly substituted with equal probability by the four substituent groups A-D shown in the figure, and the molecular weight of this molecule is 23775.1700, and the total number of isomers in this structural space shown is $6.277 \times 10^{57}$, and according to the previous formula we can learn that even if the molecule is prepared in 1 mol scale, then the resulting sample is still a single molecule mixture.

Using the hybridization method[2] shown in Fig. 14, we define Pr* as a new "heteropropyl" group obtained by equal probability hybridization of *n*-propyl (*n*-Pr) and isopropyl (*i*-Pr), and the structure of the molecule can be represented as shown in Fig. 14B. In this way, it can be seen that this "single-molecule mixture" system is equivalent to a new form of "pure substance" system. It would be a novel research work if we could synthesize this new form of substance and compare its physicochemical properties with those of the classical pure substance (e.g., the substituents in the molecular structure are all n-propyl or isopropyl).

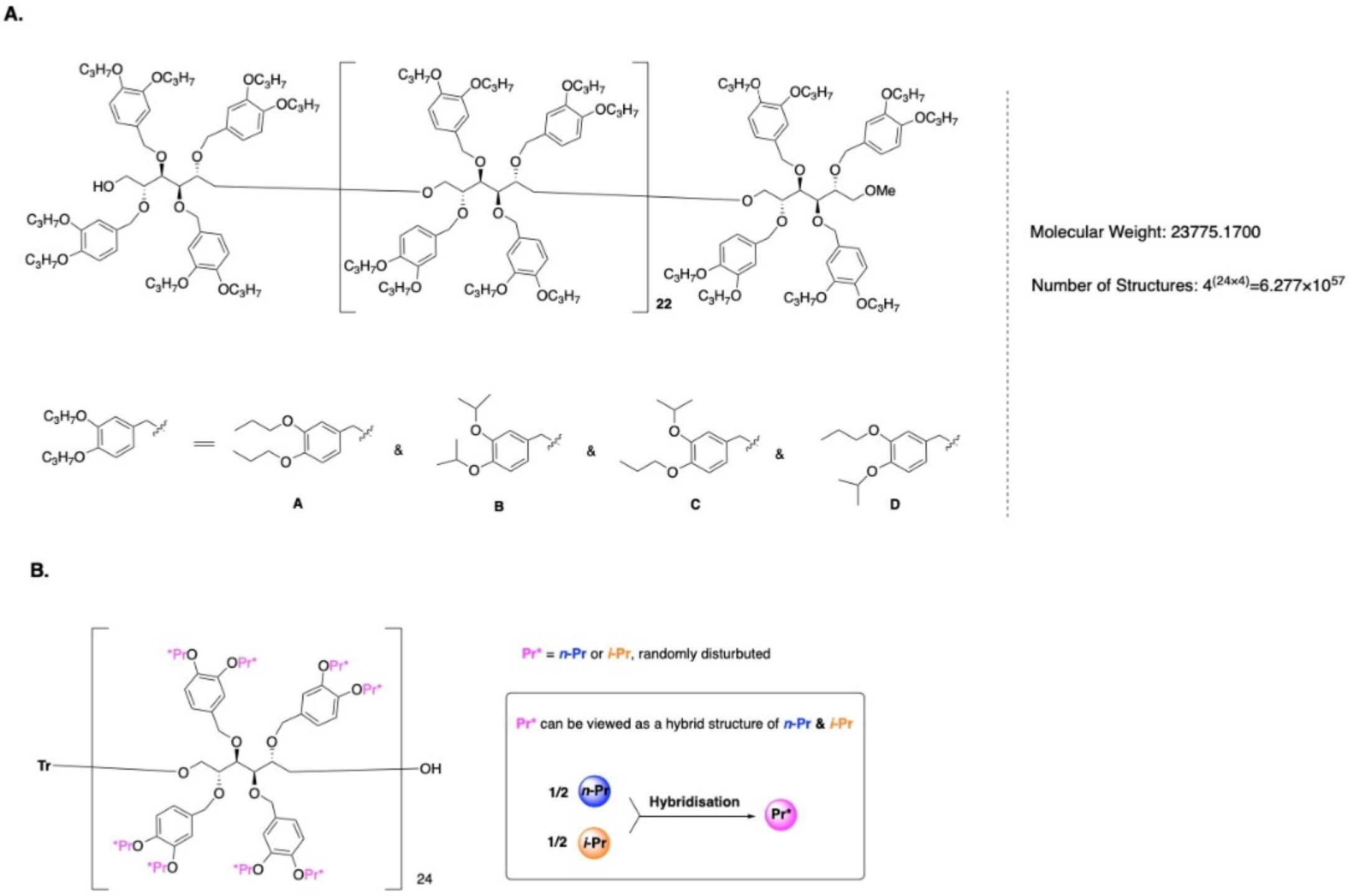

**Fig. 15**

Fortunately, according to the current development of organic chemistry, we are fully able to do this, and it can be expected that the synthesis and properties of this type of single-molecule mixture system and its comparison with the classical pure substance system will become a new research direction in the future. This research, not only has theoretical novelty, but also will lead to many surprising new discoveries!

## 10. Schematic diagrams on single-molecule mixture systems

Two schematic diagrams are particularly helpful in understanding the chemical significance and basic features of the concept of single-molecular mixture, which are briefly described here.

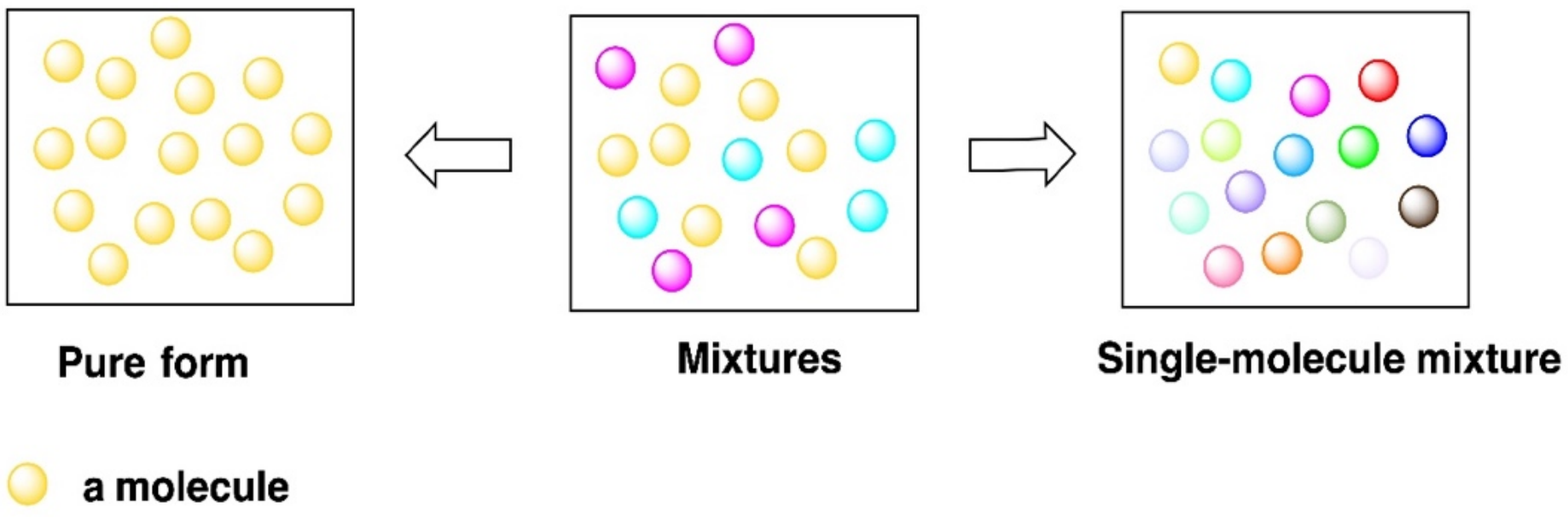

**Fig. 16**

Fig. 16 illustrates the basic chemical meaning of the concept of a single-molecular mixture. Organic matter is made up of molecules, and the absolutely pure organic matter made up of a single molecule, as shown on the left, represents an ideal "limit state". This state is very difficult to achieve. Typical organic samples are in the form of mixtures, which contain a mixture of several molecules with different structures. However, at the current level of research, no matter what, the organic samples we can prepare and obtain now always contain a large number of molecules of the same structure, as shown in the figure.

If we venture further, we can find another limit state contrary to the pure substance composed of a single molecule - a mixture of single molecules, as shown in the figure on the right. In this limit state, each molecule contained in the sample has a different structural formula and belongs to a different structure. An important reason why this state has not yet been studied is that we lack an efficient method for preparing and studying it. If such a method could be found, it would be possible to study this state of matter in detail, understand its properties, and expand its applications. Fortunately, the current level of development of theoretical and experimental techniques in organic chemistry makes it possible to find such a method, which opens up the possibility of studying single-molecule mixture systems.

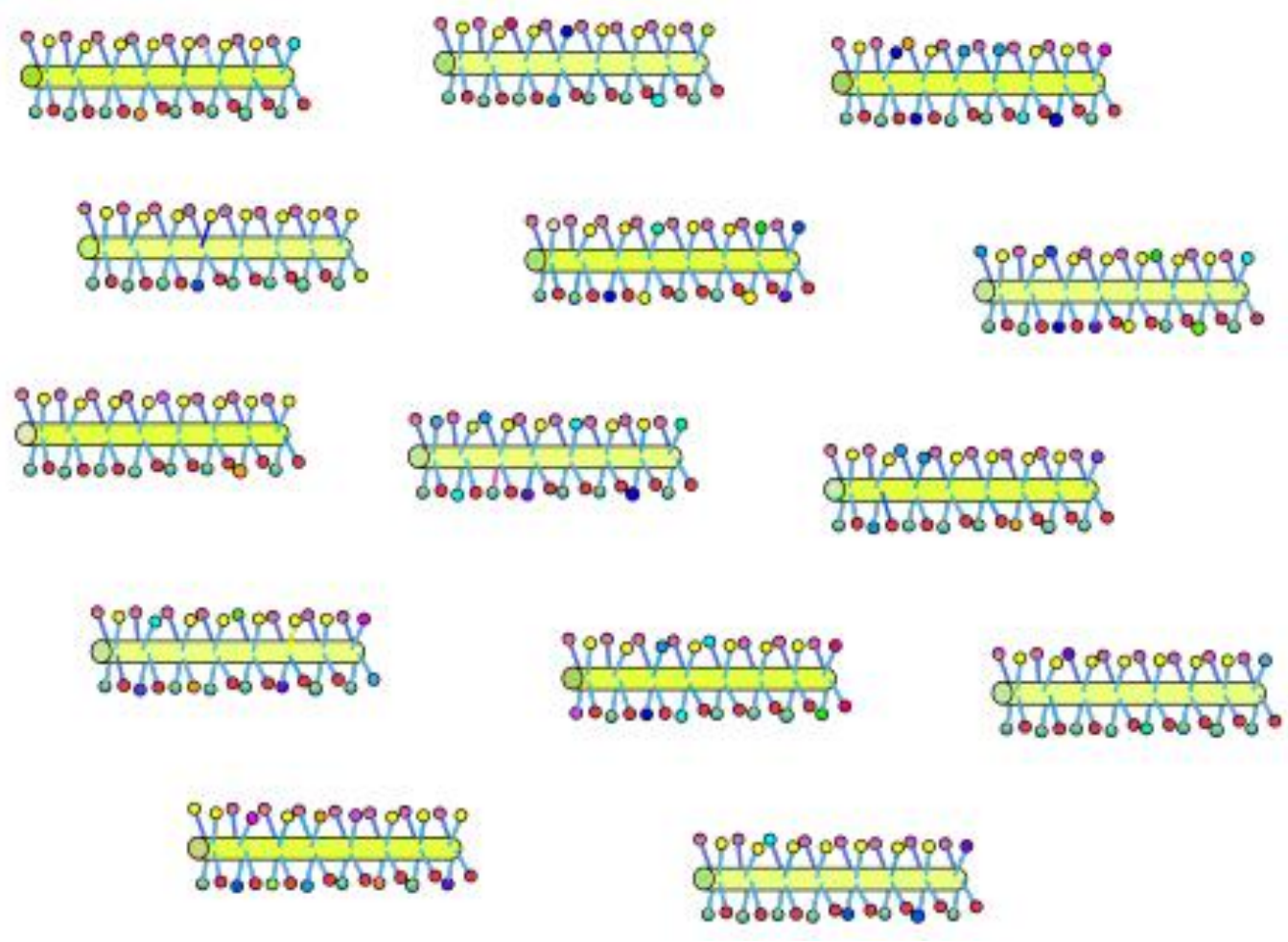

**Fig. 17**

Fig. 17 shows a schematic diagram of a typical single-molecule mixture system that we have designed. In this system, the molecules are composed of a basic skeleton and a series of substitutable side chains attached to it. Through the principle of permutation and combination, we can know that when the side chains are sufficiently numerous and randomly substituting, the possible molecular structures contained in this system are a huge number, far exceeding the actual number of molecules in the system. In this case, we can achieve the "single-molecule mixed state" with a high probability. In practical design, the type of backbone and the type of substituents can be adjusted, providing a wide range of possibilities for this research.

Such structurally similar but different states are widely found in the macroscopic real world, and it is reasonable to believe that it is also possible to realize such a state in the microscopic molecular world!

## 11. Microheterogeneity and single-molecule mixture science

**Microheterogeneity**, like **chirality**, is a fundamental property of nature, and important types of biological macromolecules, such as proteins, DNA, glycoproteins, etc., exhibit microheterogeneity[3,4], i.e., there are subtle differences in the structural composition of each molecule. This property is consistent with the fundamental properties of single-molecule mixtures of molecules discussed earlier. A single-molecular mixture is a class of organic molecules with microscopic inhomogeneities.

Chiral science, chiral synthesis, and especially asymmetric catalysis, from the conception to the practical exploration, have been greatly developed in the last half a century or so, and gradually become a very active research field in the field of modern organic chemistry[5,6]. Taking history as a reference, we can boldly foresee that microscopic inhomogeneous molecules, single-molecule mixtures science - the theory, synthetic methods, functional development of this type of molecules around the research, will gradually develop into a new energetic research direction in the field of chemistry in the 21st century.

At present, the concept of "single-molecule mixture" has been formally proposed for the first time[6]. The systematic study of this concept will lead to the formation of a new field of chemical research - single-molecule mixture science. Let's wait and see!

## 12. A method for defining the structure of microheterogeneous molecules

The structure of pure organic molecules can be defined by their chemical formulae, which are single because they are pure, and the properties of this class of molecules depend on their structure.

The situation is different for microheterogeneous molecules, which contain a huge number of different structural formulas, so how to define the structure of this class of molecules?

A reasonable solution is to define it by the structural space to which it belongs. A 'structural space' in this case consists of a basic structural skeleton and the number and type of possible substituents attached to it. It contains a large number of isomers, and the total number of isomers is determined and can be calculated precisely. The specific structural formulae of each isomer can be described one by one. Molecules from such a well-defined structural space, like different individuals of a biological 'species', have different but very similar structural formulae, and the macroscopic properties of the aggregates formed by the molecules in this space are uniform and constant. Such a 'structural space' is equivalent to a 'structural formula' in a pure substance.

In short, the concept of structural space is very important for understanding the structure of microheterogeneous molecules. Just as the structure of pure molecules is infinite, the number of structural spaces is also infinite. Constructing a structural space with a sufficiently large number of theoretical isomers, and synthesizing a small part of the molecules contained therein by random synthesis with equal probability is a basic way to prepare microheterogeneous molecules.

## 13. Synthesis of microheterogeneous molecules: a frontier topic in molecular science

**Microheterogeneity**, as a fundamental natural property, is widely presented in a range of microscopic systems such as polymer systems, biomacromolecular systems, and nanosystems [3,4,7], however, the construction of molecular systems of this form has not yet been realized at the level of organic molecules with well-defined structural compositions. Consequently, the synthesis of microheterogeneous molecules with well-defined structural compositions and determined molecular weights that exist in a single-molecule mixed state is a cutting-edge topic in the field of molecular science. If the synthesis of organic molecular systems of this form can be realized first, it will be a landmark achievement in the history of the development of the field of molecular science.

## 14. A unique feature of synthetic studies of microheterogeneous molecules

The synthetic study of microheterogeneous molecules has a unique feature compared to the usual organic synthesis studies, which are illustrated here. As shown in Figure 18, the goal of normal organic synthesis is to prepare pure molecules with the same structural composition, and in this type of study, the molecules synthesised under identical experimental conditions have the same structural formulae every time the experiment is carried out, and the parameter of 'molecular composition' can be completely reproduced. In the synthesis of microheterogeneous molecules, the molecules prepared are single-molecule mixtures with different structures for each molecule, and the molecules synthesised under identical experimental conditions have different structural formulas, and the parameter 'molecular composition' is not reproducible at all! The parameter 'composition of the molecule' is not reproducible at all!

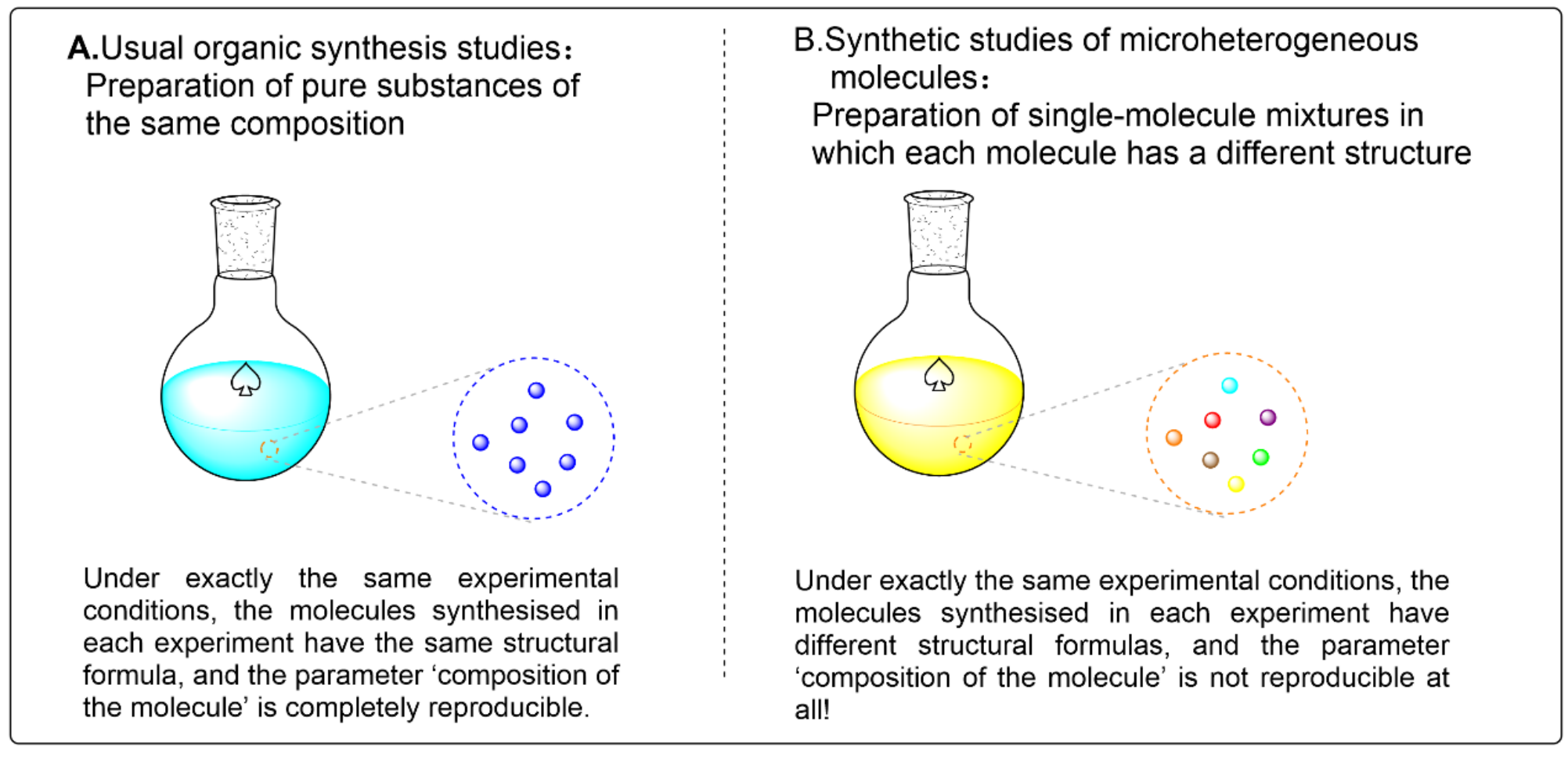

**Fig. 18**

However, this does not mean that this type of experiment is against the basic scientific principles, in fact, in many chemical studies, there is a certain amount of 'non-reproducible parameters', as illustrated here by one of the simplest crystallization experiments.

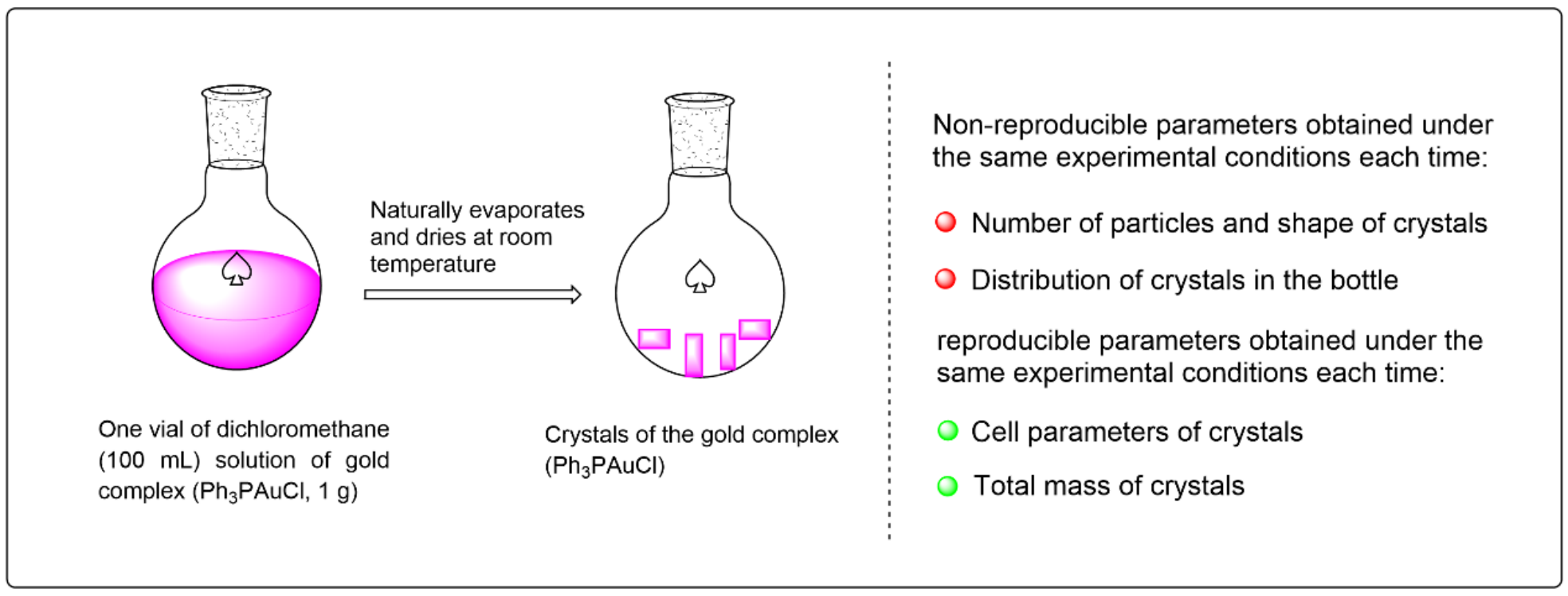

**Fig. 19**

As shown in Fig. 19, a dichloromethane solution of a certain amount of the gold complex $Ph_3PAuCl$ was prepared in a flask, and the system was naturally evaporated to dryness at room temperature to obtain crystals of the gold complex. In this experiment, the non-reproducible parameters obtained under the same experimental conditions each time included the number and shape of the particles of the crystals, and the distribution of the crystals in the bottle. Of course, there are some reproducible parameters under the same experimental conditions each time, such as the cell parameters of the crystals and the total mass of the crystals.

From this, we can find that the most unique feature of the synthesis of microheterogeneous molecules is to change the most important result parameter of synthetic chemistry, namely, the composition of the product, from being reproducible every time to being irreproducible every time, thus greatly enriching the research connotation of synthetic chemistry. By analogy with crystallization experiments, it is not difficult to find that the 'non-reproducible result parameter under the same experimental conditions every time' is in fact widespread and easy to understand. What is important is that although the result of 'the composition of molecules' is not reproducible, the macroscopic properties of molecular aggregates from the same structural space are constant and unique, just like the cell parameters in crystallization experiments.

## AUTHOR INFORMATION

Corresponding Authors

**Yu Tang —** State Key Laboratory of Chemical Biology, Center for Excellence in Molecular Synthesis, Shanghai Institute of Organic Chemistry, University of Chinese Academy of Sciences, Chinese Academy of Sciences, Shanghai 200032, China; orcid.org/0000-0002-4272-2234; Email: tangyu@sioc.ac.cn

## Notes

The authors declare no competing financial interest.

**ACKNOWLEDGMENT**
Financial support from the Youth Innovation Promotion Association of CAS (2021251) is acknowledged.